# Concurrence of ferroelectric, dielectric and magnetic behaviour in $Tb_2Ti_2O_7$ pyrochlore

B. Santhosh Kumar[1], Rajesh Narayana Perumal[2] and C.Venkateswaran[1]*

1. *Department of Nuclear Physics, University of Madras, Guindy Campus, Chennai – 600 025, India.*

2. *Department of Physics, SSN college of Engineering, Kalavakkam, Chennai 603 110, India.*

* Email: cvunom@hotmail.com

We report the ferroelectric property in $Tb_2Ti_2O_7$ pyrochlore, prepared by a two-step solid-state reaction method. A concave P-E (polarisation vs electric field) loop observed from 382 K confirms ferroelectricity, below which the loop remains as pointed or banana-shaped. Dielectric plots clearly show an anomaly at 383 K and the magnetic data reveals a deviation from 380 K supporting the observed ferroelectric ordering. The origin of ferroelectricity in $Tb_2Ti_2O_7$ is discussed which is due to the structural distortion in $TiO_6$ octahedron.

Pyrochlores are compounds that have been extensively investigated over the decades due to their fascinating spin liquid, spin ice, superconducting, metal to insulator, and magneto-resistance properties.[1-6] They have the general formula of $A_2B_2O_7$ ($A_2B_2O_6O'$) where A is a trivalent rare-earth ion like $Tb^{3+}$, $Ho^{3+}$, $Gd^{3+}$, $Pr^{3+}$, $Nd^{3+}$, $Sm^{3+}$ and B is a tetravalent transition metal ion like $Ti^{4+}$, $V^{4+}$, $Mn^{4+}$, $Ir^{4+}$, $Mo^{4+}$, $Nb^{4+}$. The A-site ion is coordinated with eight oxygen ions which can be represented as AO(6)O'(2) and B site ion is coordinated with six oxygen by forming $BO_6$ octahedra.[7] It is also interesting to note that all the six B-O bonds and six A-O(6) and other two A-O'(2) are equidistant. A-O' bond length is greater than A-O, because of the spatial repulsion of other six oxygen with A ion. Most of the pyrochlores crystallise in the *Fd-3m* space group with the crystallographic site of $A^{3+}$ ion in the position of *16d* {1/2 1/2 1/2}, $B^{4+}$ ion in *16c* {0 0 0}, O(1) in *48f* {$x$ 1/8 1/8} and O(2) in *8b* {3/8 3/8 3/8}. The only variable is O(1) in *48f* $x$ site, with $x = 0.375$ for a perfect cube and $x = 0.3125$ for a perfect octahedron about *16d*. The value of $x$ also decides the angle of octahedron - for ideal it should be 90° but ranges between 81° to 100°.[5]

$Tb_2Ti_2O_7$ is one among the pyrochlores where A and B site are occupied by $Tb^{3+}$ and $Ti^{4+}$ ions, respectively, with cubic structure (space group: *Fd-3m*). This compound continuously attracts physicists because of its intriguing properties like cooperative paramagnetism, absence of superconductivity, glassy behaviour, and absence of lattice defect or deformation[6]. S. W. Han et al., in 2004 verified the structural property of $Tb_2Ti_2O_7$ using Neutron Powder Diffraction (NPD) in the temperature range of 4.5 to 600 K and the following points were concluded from it,[8]

- It is not associated with any structural phase transition
- It forms a perfect pyrochlore lattice without any anionic disorder
- The lattice constant (a) is found to increase considerably with temperature
- For O(1) *48f* {$x$ 1/8 1/8} - $x$ decreases with temperature

Ferroelectric (FE) materials exhibit polarisation as a function of the applied electric field. Some of the pyrochlores also exhibit FE phenomenon. In 1952 W.R. Cook et al., first reported ferroelectricity in $Cd_2Nb_2O_7$ pyrochlore[9] which was followed by reports in other pyrochlores such as $Ho_2Ti_2O_7$, $Bi_2Ti_2O_7$, $Eu_2Ti_2O_7$, $Dy_2Ru_2O_7$, $RbBiNb_2O_7$ [10-14]. Table I (supplementary) shows some of the pyrochlores with structure details and coexisting properties with FE. It is interesting to note that if A site is occupied by a magnetically activated ion and if the compound shows FE property as a function of temperature (ferroelectric transition temperature), then magnetic measurements at ferroelectric transition temperature will show an anomaly as in $Ho_2Ti_2O_7$. If the A site has a non-magnetic ion and shows an FE as function of temperature, a dielectric anomaly will be seen as in $Bi_2Ti_2O_7$. Among them, Ti-based pyrochlores ($Ho_2Ti_2O_7$ and $Bi_2Ti_2O_7$) attracts our interest because, despite having cubic structure and centrosymmetry, they show FE property which is unusual. The study we report here is motivated mainly by two papers (i) Structural properties of the geometrically frustrated pyrochlore $Tb_2Ti_2O_7$ by S.W. Han et al.,[8] and (ii) Coexistence of magnetic and ferroelectric behaviours of pyrochlore $Ho_2Ti_2O_7$ by X.W. Dong et al.,.[10]

We attempt to seek FE property in $Tb_2Ti_2O_7$. This compound is not associated with any magnetic phase transition from 20 to 800 K. This leads to the confirmation if $Tb_2Ti_2O_7$ shows any sign of FE it should also mediate the dielectric property. We report the ferroelectric property in $Tb_2Ti_2O_7$ from polarisation versus electric field above 382 K, and the dielectric and magnetic measurements also show an anomaly around 383 K evincing FE probably due to a distortion in $TiO_6$ octahedron at T > 382 K. As far as we know there are no experimental reports of ferroelectricity in $Tb_2Ti_2O_7$.

Polycrystalline $Tb_2Ti_2O_7$ was synthesized by firing the high energy ball milled powder composition of $Tb_4O_7$ and $TiO_2$, as discussed elsewhere, and the crystalline nature was verified using X-Ray diffraction.[15] For ferroelectric measurement a pellet of 8 mm diameter and 1 mm thickness sintered at 500 °C for 5h (to remove moisture) had been used. A thin layer of high quality silver paste was coated on both sides of the pellet and the polarization versus electric field measurements was done at 20 Hz using Marine high-end PE loop ferroelectric test system 20PE-1 kHz-1N. The dielectric property had been studied using a two probe set up with platinum electrodes sandwiching a pellet of 8 mm diameter and 0.95 mm thickness (also sintered at 500 °C for 5 h to remove the moisture content in it) and connected to the Solatron 1260-impedance analyser. The magnetisation measurement was performed in a Lakeshore 7410 Model Vibrating sample magnetometer.

Figure 1 (a to f) shows the polarisation vs electric field loop (PE) in the temperature range of 301 K to 382 K. The plot till 371 K appears to be pointed or banana type of a loop.[16] The maximum polarisation value is also nearly the same in the range of $1.5\mu C/cm^2$ to $1.4\mu C/cm^2$ at 6 kV/cm and the ferroelectric property is absent in the prepared $Tb_2Ti_2O_7$ pyrochlore till 371 K. The plot 1.f shows the PE loop taken at 382 K. The loop appears to be concave and the maximum polarisation value decreases evincing the ferroelectric behaviour from 382 K. The PE loop from 391 K to 412 K (figure 2) also clearly shows the hysteresis behaviour confirming the ferroelectric nature. The maximum polarisation decreases from 0.85 µC/cm$^2$ to 0.28 µC/cm$^2$ in the temperature range of 382 K to 412 K. Also, the magnitude of polarization observed in $Tb_2Ti_2O_7$ is comparable with $Ho_2Ti_2O_7$.[10] It is to be noted that temperature-dependent FE property is reported in some of the pyrochlores like $Bi_2Ti_2O_7$ (at 308 K),[17-18] $Ho_2Ti_2O_7$ (at 60 K),[10] and $Cd_2Nb_2O_7$ (at 77 K).[9]

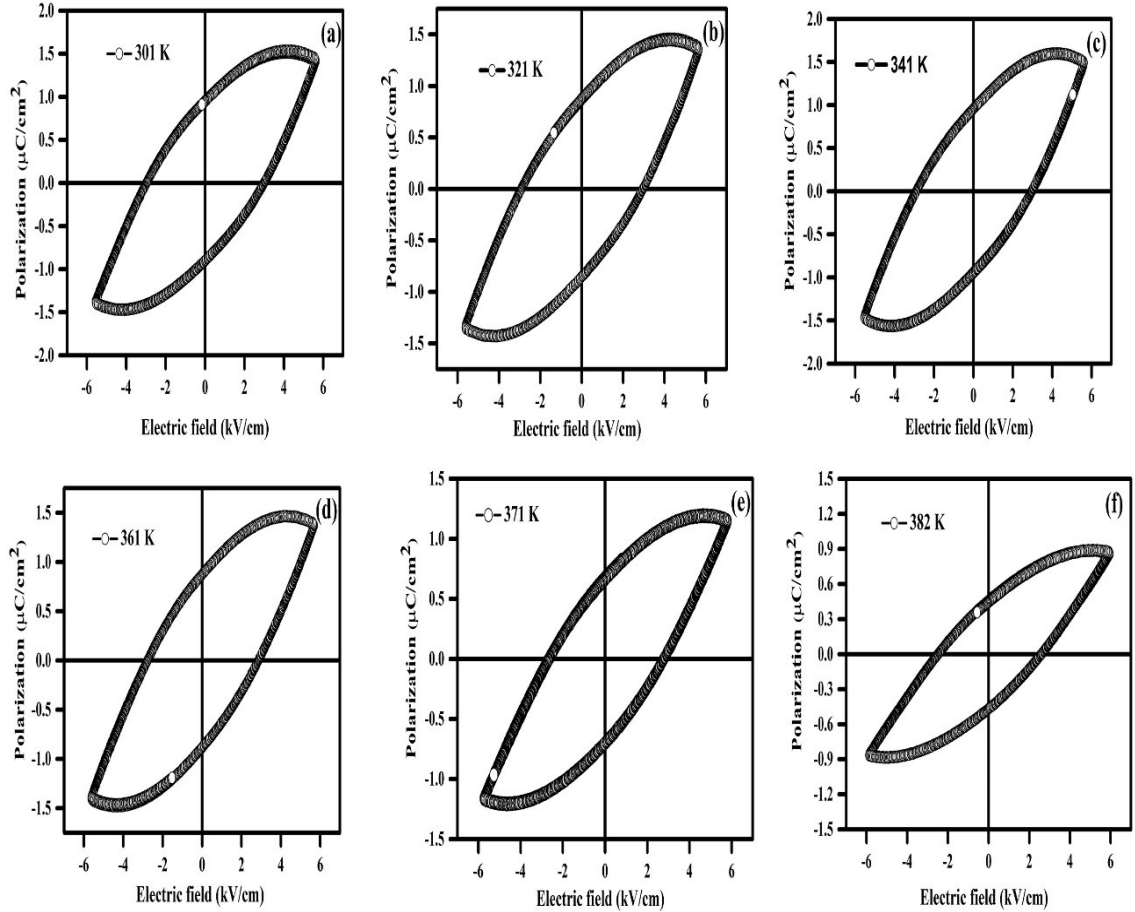

Figure 1: Polarisation vs electric field (hysteresis) loop for $Tb_2Ti_2O_7$ at various temperatures, 'a - e' the loop appears to be pointed from 301 K to 371 K and (f) the loops clearly show a change in trend with the concave region from 382 K.

Figure 3(a) shows the real part of the dielectric constant (ε') of $Tb_2Ti_2O_7$ from 303 K to 440 K. The ε' at room temperature is observed as 40.8 which is in agreement with the reported value[19]. The real part of dielectric constant decreases as the temperature increases and remains constant from 383 K – which is a sign of FE in the prepared sample. To confirm the behaviour, 1/ ε' vs log *f* is plotted that will reveal the transition clearly as a function of temperature (fig 3.b). A dielectric anomaly seen at 383 K is the sign of ferroelectricity. Figure 3 (c) and (d) shows the imaginary part of impedance (Z") and the inverse electric modulus (M") vs *f*. Z" plot (Fig 3.c) attains a maximum value at 383 K and starts decreasing beyond this temperature, and 1/M" plot confirms the observed FE.

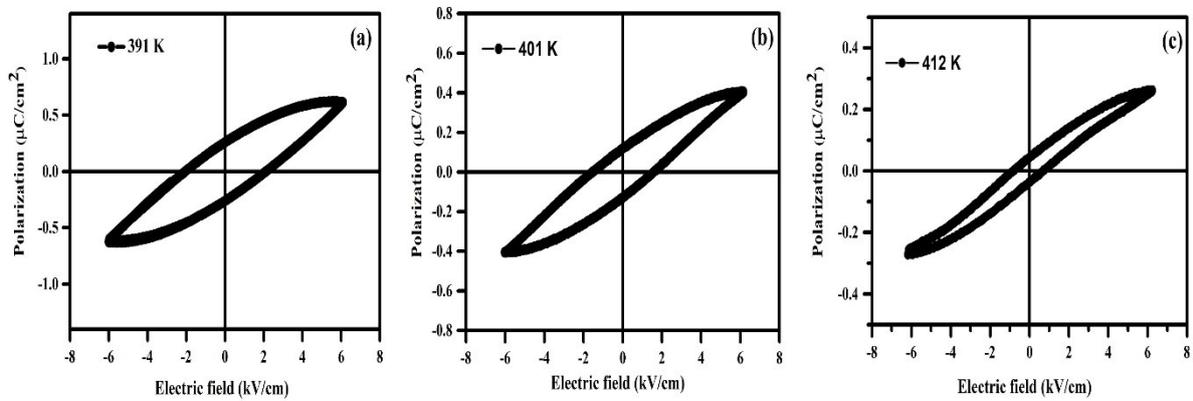

Figure 2: PE-Loop from 391 K to 412 K confirming ferroelectric property in $Tb_2Ti_2O_7$ above 382 K.

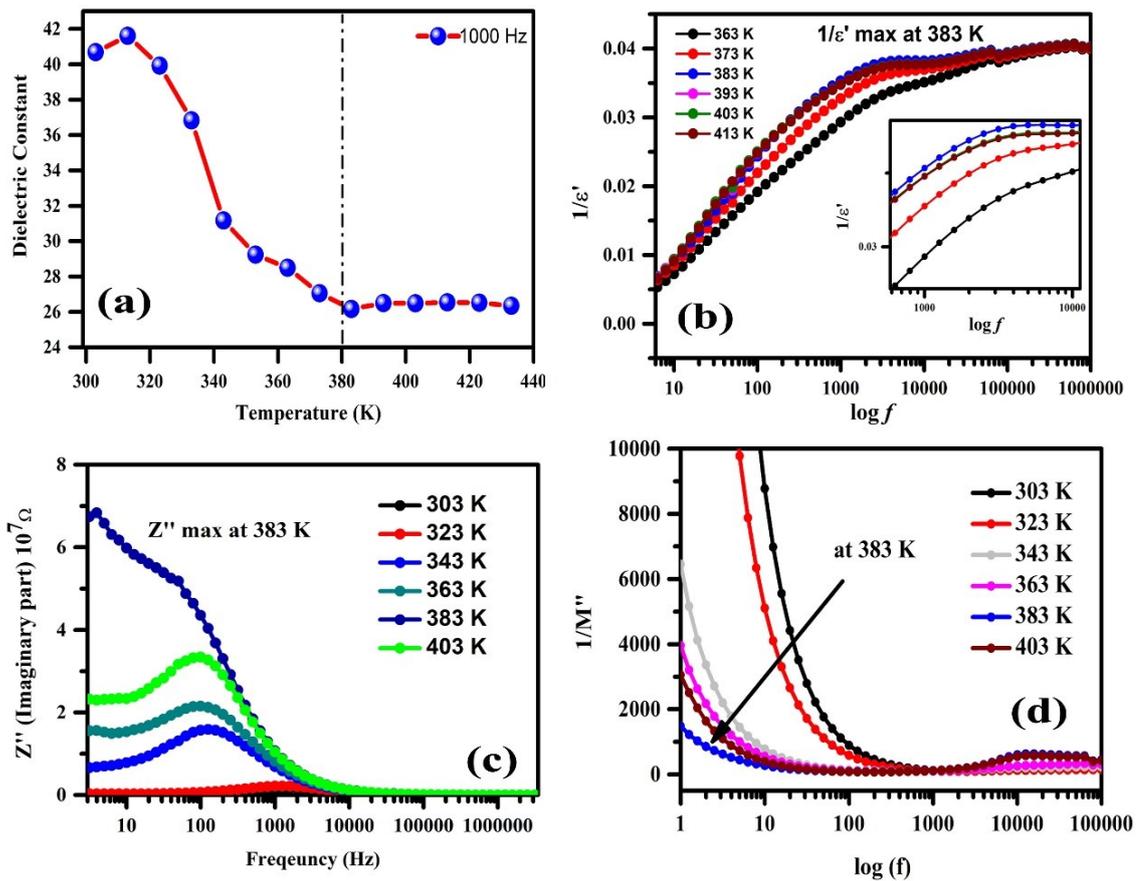

Figure 3: (a) Real part of dielectric constant ($\varepsilon'$) as a function of temperature from 303 K to 433 K at 1000 Hz for $Tb_2Ti_2O_7$, (b) $1/\varepsilon'$- to compare the response of material at 383 K, and the plot shows a significant increase at 383. Inset shows a maximum peak between 1kHz to 10 kHz (C). Imaginary part of impedance (Z") shows a maximum value at 383 K beyond which a decreasing trend is seen and (d) Imaginary part of electric modulus (1/M") vs frequency at various temperature shows a dielectric anomaly, and a ferroelectric loop is observed.

The magnetic moment as a function of temperature shows a deviation from 380 K (figure 4.a) which supports the possible ferroelectric nature of $Tb_2Ti_2O_7$, the origin of which is discussed based on the crystallographic features. As discussed in the beginning, even a small shift in the ideal position of O (1) - 48f – {$x$ 1/8 1/8} will result in the distortion of octahedron altering the bond angle of $O^{2-}$-$Ti^{4+}$-$O^{2-}$. In other words, symmetry breaking without any structural phase transition[10] drives the material to the ferroelectric state. The above statement is also true in the case of other pyrochlores that are driven to the FE state.[8, 20] The variation in the coordinate(s) of O(1) - 48f – {$x$ 1/8 1/8} had been reported by S.W. Han et al., for $Tb_2Ti_2O_7$ where $x$ decreases with increase in temperature.

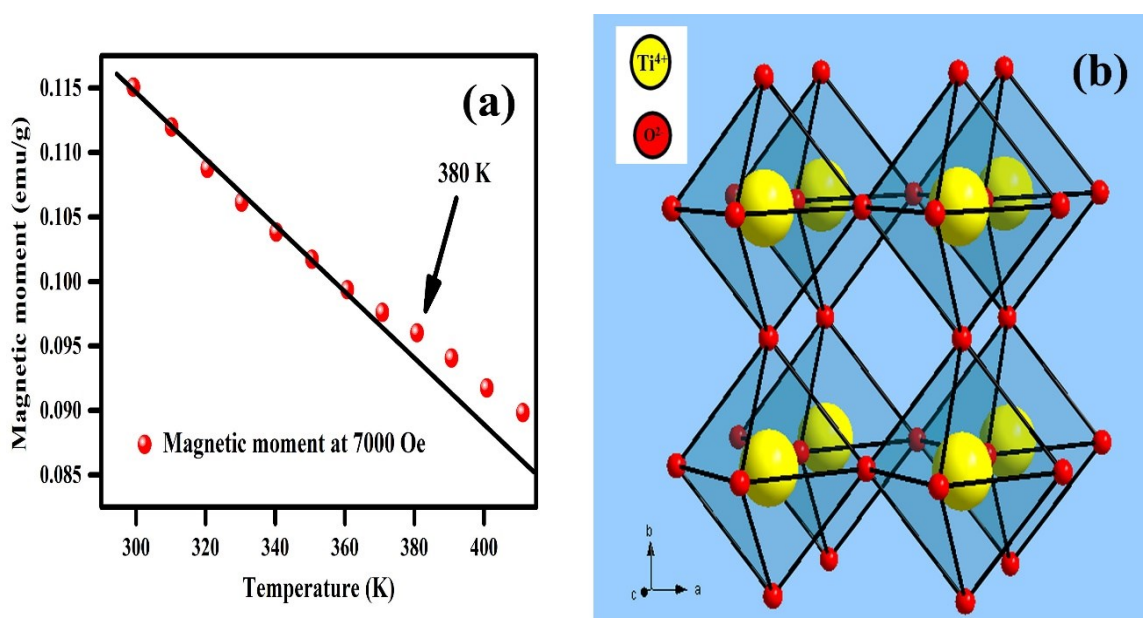

Figure 4: (a) shows the plot of magnetic moment vs temperature at 7000 Oe. A deviation from 380 K is seen that supports the ferroelectric nature of $Tb_2Ti_2O_7$ and (b) shows the arrangement of $TiO_6$ octahedron in $Tb_2Ti_2O_7$ crystal structure - the yellow spheres indicate $Ti^{4+}$ and red spheres indicate $O^{2-}$ ions.

Therefore, beyond 382 K the alteration in $x$ of O(1) is significant, contributing to the distortion and the net polarization. Figure 4.b shows the ideal $TiO_6$ octahedra in $Tb_2Ti_2O_7$. It can be viewed easily that even a small shift in O (1) can drive the material into the ferroelectric phase as reported in the compounds $Ho_2Ti_2O_7$, $Bi_2Ti_2O_7$, $Eu_2Ti_2O_7$.[10, 12, 17-18] Ferroelectricity in $Bi_2Ti_2O_7$ at 308 K (35 °C) is also due to structural distortion in $TiO_6$ octahedron.[18] But, the ferroelectric ordering is observed at a comparatively lower temperature due to the presence of lone pair in $Bi^{3+}$ cation that easily distorts the $TiO_6$ octahedra and leading to the ferroelectric state at a lower temperature (compared with $Tb_2Ti_2O_7$ at 382 K). Absence of lone pair electron in $Tb^{3+}$ makes the compound to be ferroelectric at a much higher temperature (than $Bi_2Ti_2O_7$). However, in addition to the structural distortion, any other possible reason for ferroelectric property in Ti-based pyrochlores need to be understood.

In summary, the letter detailing the ferroelectric property in cubic $Tb_2Ti_2O_7$ from 382 K is affirmed due to the structural distortion in $TiO_6$ octahedron. The dielectric constant, imaginary part of the impedance and electric modulus shows a variation around 382 K, supported by the deviation in the magnetic moment from 380 K. The coexistence of ferroelectricity with dielectric and magnetic property is thus established, beyond 382 K.

**Acknowledgement**

BSK thanks Department of Science & Technology, Ministry of Science and Technology (DST) for the award of DST Inspire fellowship (SRF-IF-140582). BSK acknowledges Mr Arulmani Marimuthu, SSN College of Engineering, Chennai, for ferroelectric measurements, and Sophisticated Analytical Instruments Facility (SAIF), IIT-Madras for magnetic measurement.

**Data availability statement**

The data that support the findings of this study are available from the corresponding author upon reasonable request.

**Conflict of interest**

The authors declare, there is no conflict of interest.

**Supplementary:**

*Table 1: Shows some of the pyrochlores which exhibit Ferro electricity and coexisting properties, with their space group and symmetry indicated.*

| Pyrochlore Compound | Structure | Space Group | Symmetry | Coexisting property with FE | Ref. |
|---|---|---|---|---|---|
| $Ho_2Ti_2O_7$ | Cubic | *Fd-3m* | Centro | Magnetic | 1 |
| $Bi_2Ti_2O_7$ | Cubic | *Fd-3m* | Centro | Magnetic | 2-4 |
| $Eu_2Ti_2O_7$ | Cubic | *Fd-3m* | Centro | Electric | 5 |
| $Dy_2Ru_2O_7$ | Cubic | *Fd-3m* | Centro | Magnetic | 6 |
| $Cd_2Nb_2O_7$ | Cubic | *Fd-3m* | Centro | Dielectric / electric | 7 |
| $RbBiNb_2O_7$ | Orthorhombic | *P2$_1$am* | Non-centro | Dielectric / Electric | 8 |